\documentclass[letter]{aa}

\usepackage{amsmath, amssymb, amsfonts, amscd}
\usepackage{mathtools} 
\usepackage{array} 
\usepackage[varg]{txfonts} 
\usepackage{bm}

\usepackage{graphicx} 

\usepackage{natbib}
\bibpunct{(}{)}{;}{a}{}{,} 

\begin{document}

\title{Understanding hydrodynamical wave-driven shear mixing\\ in stellar radiation zones}
\subtitle{Looking in the mirror of the dyapicnal oceanic mixing
}

\author{S. Mathis\inst{1}}

 \institute{Universit\'e Paris-Saclay, Universit\'e Paris Cit\'e, CEA, CNRS, AIM, F-91191 Gif-sur-Yvette, France\\
 \email{stephane.mathis@cea.fr}}

\date{Received ... / accepted ...}

\abstract 
{Stellar radiation zones play a key role in the long-term magneto-rotational and chemical evolution of stars and of their planetary and galactic neighborhood. As parts of the oceans and of the atmosphere of the Earth, they are stably stratified and rotating. Therefore, their dynamics is controlled by the Archimedean buoyancy force and the Coriolis acceleration. Asteroseismology and high-resolution spectroscopy have demonstrated that they are the seat of an efficient extraction of angular momentum and of a mild mixing of chemicals that must be understood. In this context, particle tracing in recent nonlinear hydrodynamical equatorial numerical simulations of stellar radiation zones where internal gravity waves (hereafter IGWs) are propagating led to the measurement of an effective diffusivity following the prescriptions derived by Garcia-Lopez \& Spruit and by Zahn for the inflectional instability of the vertical shear of low-frequency IGWs. However, the associated instability criteria are not fullfiled. This effective diffusivity is found to scale as the squared velocity of IGWs for every rotation rates. Other dependences have also been derived in the literature, for instance in the case of the Stokes displacement.}
{We aim to provide a physical interpretation of these results.}
{To reach this objective, we propose to explore the parameterisation for the mixing of particles, which has been proposed in another rotating stably stratified systems: the oceans of the Earth. A foundation stone in physical oceanography is the so-called Osborn \& Cox energetic balance that leads to an effective dyapicnal diffusivity for the transport of matter that scales as the ratio of the dissipation of the fluctuating flows over the squared Brunt-V\"ais\"al\"a stratification frequency. We apply this parameterisation for the mixing to the field of low-frequency IGWs propagating in stellar radiation zones.}
{We demonstrate that the effective dyapicnal diffusivity, which is widely used in the modeling of the oceanic general circulation, is equivalent to the eddy diffusivity derived by Zahn for the inflectional instability of the vertical shear applied to low-frequency IGWs. This allows us to characterize the corresponding energetic balance where the power extracted by the waves from the mean flows is balanced by their dissipation and by the power produced by their buoyancy flux for any rotation rate. This bridge established between results obtained in geophysical and stellar fluid dynamics illustrates the high interest of a cross-fertilisation between these research fields when aiming to evaluate and model robustly the effects of small-scale and short time scales phenomena such as IGWs on the long-term evolution of the large-scale oceanic, atmospheric and stellar general circulation.} 
{}

\keywords{Waves, stars: rotation, stars: evolution, Planets and satellites: atmospheres, oceans, methods: analytical}

\titlerunning{}
\authorrunning{Mathis}

\maketitle


\section{Introduction}

In a standard vision of stellar structure and evolution, stably stratified stellar radiation zones are assumed to be in hydrostatic and radiative balances while being only the seat of microscopic mixing processes. However, space-based seismology of our Sun and other stars have revealed that they are the seat of a strong transport of angular momentum and of a mild macroscopic mixing of chemicals \citep[e.g.][]{Pinsonneault1997,AMR2019,JCD2021,AT2024}. These mechanisms have a strong impact on the evolution of stars, for instance on their age and on their chemical properties, which impact at largest scale the galactic evolution \citep{Maeder2009}. Several mechanisms have been identified as serious candidates to reproduce simultaneously the rotation probed by helio- and asteroseismology at different depths in our Sun and stars, respectively, and their chemical composition. First, the hydrodynamical instabilities of the differential rotation have been studied \citep{Zahn1983,PratLignieres2014,Parketal2021,Dymottetal2023,Garaud2024,ParkMathis2025} in combination with the meridional circulation they trigger with angular momentum losses at the stellar surface \citep{Zahn1992,MaederZahn1998,MZ2004}. This so-called "Type-I" rotational mixing has obtained successes in explaining the observed chemical abundances in massive stars \citep{MaederMeynet2000} but has failed to explain the strong extraction of angular momentum observed in the Sun and in stars of different masses at different evolutionary stages \citep{TCetal2010,Marquesetal2013,Ouazzanietal2019}. Next, magnetic fields and their instabilities \citep{Spruit1999,Spruit2002,MathisZahn2005,Barrereetal2026} and low-frequency IGWs excited by the turbulence at the interfaces with adjacent convective regions and in their bulk \citep[e.g.][]{Press1981,Schatzman1993,Zahnetal1997,Fulleretal2014,Rogers2015,Pinconetal2017} have been examined; this is the so-called "Type-II" rotational mixing. Both mechanisms have demonstrated their potential ability to reproduce solar and stellar rotational and chemical properties (we refer the reader to \cite{Eggenbergeretal2005} and \cite{Eggenbergeretal2002} in the case of magnetic fields and \cite{CT2005} and \cite{TC2005} in the case of IGWs). Particularly, IGWs can be an efficient and discriminant vector of chemical mixing of light elements in late-type stars \citep{MontalbanSchatzman2000,CT2005}, in evolved stars \citep{DT2003}, and in early-type stars \citep{Herwigetal2023,Mombargetal2025}.\\

In this context, \cite{RME2017} computed nonlinear hydrodynamical equatorial simulations of a 3$M_{\odot}$ star, which simultaneously computes the dynamics of its convective core and its radiative envelope. Using a particle tracer method, they show that IGWs act as an effective diffusion for the concentration of chemicals and that the associated eddy-diffusion scales as the squared velocity of IGWs. These results have been confirmed in a subsequent series of articles \citep{Vargheseetal2023,Vargheseetal2024} where different stellar masses, evolutionary stages and rotation have been examined. If these results have been debated by \cite{Mortonetal2025}, they have been consolidated and confirmed by \cite{Vargheseetal2025} who demonstrated that the functional dependence of the diffusion coefficient as a function of the velocity of IGWs can be explained by their own vertical shear instabilities. However, two questions remain open. On the one hand, IGWs observed in the work of \cite{Vargheseetal2025} do not fullfill the vertical shear instability criterion. On the other hand, other functional dependences of the diffusion have been proposed in the literature. For instance, it scales as the fourth power of the velocity of IGWs if the diffusion is due to the so-called Stokes displacement induced by the thermal dissipation of IGWs.\\ 

\begin{figure}[h!]
\centering
\includegraphics[width=0.475\textwidth]{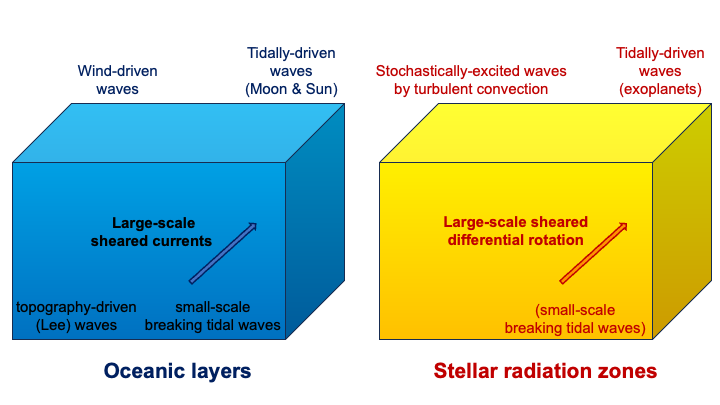}        
\caption{Analogies between stably stratified rotating oceanic layers and stellar (planetary) radiation zones. In both cases, internal gravity and gravito-inertial waves are excited in the adjacent convective layers, by the topography in the case of the Earth, and by tides exerted by the Sun and the Moon in the case of the Earth and those induced by planetary or stellar companions in the astrophysical context.}
\label{Fig1}
\end{figure}

To try to adress these two questions, we propose in this work to exploit the similarities of stellar radiation zones with the oceans. Both systems are the seat of a complex field of propagative IGWs excited at their boundaries: by adjacent turbulent convective regions and potential planetary companions in stars \citep[e.g.][]{Zahn1975,BarkerOgilvie2010,AMA2021} and by the winds, the topography and the solar and lunar tides in the oceans; this is illustrated in Fig. \ref{Fig1}. In the oceanic case, an important litterature has been developed since the seminal work by \cite{OsbornCox1972} and \cite{Osborn1980} to evaluate the mixing triggered by small scale and short time scale fluctuating flows, possibly including progressive IGWs, and the related effective diffusivity, which is used as a parameterisation in oceanic global circulation models \citep[e.g.][and references therein]{MacKinnonetal2017}. We recall in Sec. \ref{sec:OSCM} the physical basis of the Osborn \& Cox model and the related diffusion parameterisation. We show in Sec. \ref{sec:SSE} how it can be applied to the stellar case allowing us to provide a complementary understanding of the results that have been obtained recently in the literature. Finally, we discuss in Sec. \ref{sec:DiscConclu} the future developments and we present our conclusions.

\section{The Osborn-Cox closure model}
\label{sec:OSCM}   

In their work, Osborn \& Cox (1972) and Osborn (1980) have proposed a closure model to parameterise and evaluate the mixing in rotating stably stratified oceanic layers. They split the components of the velocity and thermodynamical quantities as:
\begin{equation}
u_{i}={\overline u}_{i}+u_{i}^{'}\quad\hbox{and}\quad X={\overline X}+X^{'},
\end{equation}
where ${\overline{\cdot\!\cdot\!\cdot}}$ and $\cdot\!\cdot\!\cdot^{'}$ are their mean and fluctuating components, respectively. Next, they consider the Reynolds equation for the fluctuating components: 
\begin{eqnarray}
\lefteqn{\frac{D{\overline e}}{D t}=-\frac{\partial}{\partial{x_i}}\left({\overline{p^{'}u_{i}^{'}}}\right)-\frac{\partial}{\partial{x_i}}\left(\overline{e\,u_{i}^{'}}\right)+\frac{\partial}{\partial x_{j}}\left(2\,{\rho}_{B}\nu\overline{s_{ij}\,u_{i}^{'}}\right)}\nonumber\\
&&-\overline{\rho^{'}u_{r}^{'}}{\overline g}-{\rho}_{B}\,\epsilon-{\rho}_{B}\,\overline{u_{i}^{'}u_{j}^{'}}\,\frac{\partial{\overline U}_{i}}{\partial x_{j}},\nonumber\\
&&\hbox{where}\quad
{e}=\frac{1}{2}{\rho}_{B}\,{u_{i}^{'}u_{i}^{'}}\quad\hbox{and}\quad
\epsilon=2\,\nu\,{\overline{s_{ij}s_{ij}}}
\label{eq:Reynolds}
\end{eqnarray}
are the mean kinetic energy of the fluctuating components of the velocity and their viscous energy dissipation with the viscosity $\nu$, respectively. We consider fluctuating velocities and perturbations that vary over small characteristic length-scales when compared to those of the variations of the hydrostatic background. This allows us to assume the Boussinesq approximation and we introduce ${\rho}_{B}$ the corresponding constant and uniform reference density. We have introduced the Lagrangian derivative $D_{t}\equiv\partial_{t}+u_{i}^{'}\partial_{x_{i}}$ and the stress tensor
\begin{equation}
s_{ij}=\frac{1}{2}\left(\frac{\partial u_{i}^{'}}{\partial x_j}+\frac{\partial u_{j}^{'}}{\partial x_i}\right).
\end{equation}
When assuming a statistical steady state and that the first three redistribution flux terms on the right-hand side of Eq. (\ref{eq:Reynolds}) can be neglected in front of the three last terms, \cite{OsbornCox1972} and \cite{Osborn1980} identified the balance between the power transfered from the mean flow to the fluctuating velocities, their viscous dissipation and the power of their buoyancy flux:
\begin{equation}
\overline{u_{i}^{'}u_{j}^{'}}\frac{\partial{\overline U}_{i}}{\partial_{x_j}}=-\epsilon-\frac{\overline{\rho^{'}u_{r}^{'}}{\overline g}}{\rho_{B}}.
\label{OCBalance}
\end{equation}
Next, they evaluate the transport of mass using an effective diapycinal eddy diffusivity and the flux Richardson number that they defined as:
\begin{equation}
D_{\rho}=\frac{\overline{\rho^{'}u_{r}^{'}}{\overline g}}{\rho_{B}\,N^{2}}
\quad\hbox{and}\quad
R_{f}=-\frac{\overline{\rho^{'}u_{r}^{'}}{\overline g}}{\rho_{B}\,\overline{u_{r}^{'}u_{h}^{'}}\partial_{r}{\overline U}_{h}}.
\end{equation}
This latter measures the fraction of fluctuating (turbulent) power generated by the mean shear that is converted into mixing against the stabilising stratification term. They follow from Eq. (\ref{OCBalance}):
\begin{equation}
D_{\rho}=\Gamma_{\rho}\frac{\epsilon}{N^{2}},
\quad
\hbox{where}
\quad
\Gamma_{\rho}=\frac{R_{f}}{\left(1-R_{f}\right)}
\end{equation}
is an efficiency parameter associated with possible mixing processes that increases with $R_{f}$. Following the path proposed by Osborn \& Cox, \cite{Hamiltonetal1989} demonstrated that this relationship can also be applied to other fluctuating thermodynamical quantities: 
\begin{equation}
D_{X}=\Gamma_{X}\frac{\epsilon}{N^{2}},
\label{Eq:OCP}
\end{equation}
where $X\equiv\left\{T,S\right\}$ with $T$ and $S$ the temperature and the oceanic salinity (the mean molecular weight in an astrophysical body), respectively \citep[we refer the reader to][for the detailed derivation of the efficiency parameters $\Gamma_{X}$ as a function of $\Gamma_{\rho}$ and of thermodynamical quantities]{Hamiltonetal1989}.

This parameterisation has been widely used in Oceanic General Circulation Models \citep{MacKinnonetal2017}. Since this parameterisation aimed to describe the mixing of chemicals in oceanic rotating stably stratified layers, it is of high interest to apply it in the astrophysical context of stellar (and planetary) rotating stably stratified radiation zones.\\

\section{Recovering the prescription for the wave vertical shear-driven mixing in stellar radiation zones}
\label{sec:SSE}

\subsection{The effective diffusivity}

We consider a (rotating) stably stratified stellar radiation zone. In such a region, the buoyancy restoring force is limiting the motion along the radial direction of the entropy and chemical stratifications. In the case of low-frequency internal (gravito-inertial) gravity waves, which are efficient to transport momentum and matter \citep[e.g.][]{Press1981,Schatzman1993,Zahnetal1997,Mathisetal2008}, this leads to $u_{r}\!<\!\!<u_{h}$, where $u_{r}$ and $u_{h}$ are the radial and horizontal components of their velocity, respectively. Considering the Osborn-Cox prescription given in Eq. (\ref{Eq:OCP}), we thus obtain that the energy dissipation can be approximated by:
\begin{equation}
\epsilon\sim \nu \left(\partial_r u_h\right)^2.
\end{equation}
Assuming that the fluctuating motions, which are considered in the Osborn-Cox modelling, can be low-frequency internal gravito-inertial or gravity waves, we thus obtain for the eddy-like diffusion coefficient of a passive scalar the following form:
\begin{equation}
D_{\rm X}\sim \Gamma_{X}\frac{\epsilon}{N^2} \sim \Gamma_{X}\frac{\nu \left(\partial_r u_h\right)^2}{N^2} \sim {\widetilde \Gamma}_{X}\frac{K\left(\partial_r u_h\right)^2}{N^2},
\end{equation} 
where we have introduced ${\widetilde\Gamma}_{X}=\Gamma_{X}\,P_{r}$ with $P_{r}=\nu/K$ the Prandtl number, which evaluates the relative strength of the viscosity and of the heat diffusivity $K$ (in stellar radiation zones $P_{r}<1$ while $P_{r}\sim7$ in oceans). This straightforward result is of high interest since we recover the \cite{Zahn1992} prescription applied to the vertical shear of low-frequency internal waves \citep{Vargheseetal2025}, which has been first derived by \cite{GLS1991}. Recently, \cite{Vargheseetal2025} have robustly demonstrated numerically that the transport of chemicals by internal waves they observed in their series of nonlinear hydrodynamical equatorial numerical simulations of non-rotating and rotating early-type and late-type stars can be modeled with such a functional dependence.

If we use again that low-frequency internal waves are rapidly oscillating along the radial direction and that these dependence can be described using the JWKB asymptotic approximation where $\vert\partial_r u_h\vert\sim k_r \vert u_h\vert$, with $k_{r}$ being the radial component of the wave vector, we identify that $D_{X} \propto \vert u_h\vert^2$. This dependence has been first identified by \cite{RME2017} in their similar numerical simulations of a $3\,M_{\odot}$ star. Next, it has been confirmed by \cite{Vargheseetal2023}, who have studied different stellar masses and evolutionary stages, and by \cite{Vargheseetal2024}, who have examined the dependence as a function of stellar rotation. In this regime, the vertical dependence of the horizontal component of the velocity of the waves can be expressed using the JWKB approximation \citep[e.g.][]{Mathis2025} and we get:
\begin{eqnarray}
\lefteqn{{u}_{h}\left(r,X_h,t\right)={\widehat u}_{h}\left(r_0,\omega\right)\times}\nonumber\\
&&\left(\frac{\epsilon\left(r\right)}{\epsilon\left(r_0\right)}\right){\mathcal H}\left(X_{h}\right)\exp\left[i\,\left(\pm \int_{r_0}^{r}k_r\left(r'\right){\rm d r'}+\omega t\right)\right],
\end{eqnarray}
where $r_0$ is the radius of their excitation region, $\epsilon\left(r\right)$ theirs slowly varying JWKB envelope that can include their linear damping, $X_h$ a generic horizontal coordinate adapted to the studied geometry and ${\mathcal H}$ the corresponding function describing the horizontal variations of the waves (we refer the reader to the Appendix \ref{Appen:A} for their explicit expressions in the different used geometries), $\omega$ their angular frequency, and $t$ the time. The effective diffusivity becomes:
\begin{equation}
D_{\rm X}\!=\!\frac{1}{2}{\widetilde\Gamma}_{X}\frac{K}{N^2}\left[k_r\left(r\right)\frac{\epsilon\left(r\right)}{\epsilon\left(r_0\right)}{\widehat u}_{h}\left(r_0,\omega\right)\right]^2\!\!\left<{\mathcal H}\left(X_{h}\right)\left[{\mathcal H}\left(X_{h}\right)\right]^{*}\right>_{h},
\label{Eq:Dstar}
\end{equation}  
where $\left<\cdot\!\cdot\!\cdot\right>_{h}$ is the average over the horizontal directions in the considered geometry and $^{*}$ the complex conjugate.\\

A key question at this point is to know if low-frequency internal waves can legitimately be considered as the fluctuating field in the Osborn-Cox methodology when we apply it to stellar radiation zones. We think that a positive answer can be assumed. On the one hand, low-frequency internal waves have very short radial wavelengths when compared to the characteristic lengths of variation of the hydrostatic stellar gravity and thermodynamical quantities (the density, the entropy, the temperature, and the mean molecular weight) and of the differential rotation. On the other hand, low-frequency internal waves have characteristic periods that are very short when compared to the characteristic times of the secular extraction/deposit of angular momentum they trigger \citep[we refer the reader to the detailed discussion in][]{TC2005}. To assume such an hypothesis, we have assumed that the interactions occurring between the waves and the differential rotation (and the associated mean zonal velocity) such as the so-called quasi-biennal oscillation observed in the equatorial region of the terrestrial atmosphere \citep[e.g.][]{Baldwinetal2001} and its stellar analog, the so-called shear layer oscillation \citep[e.g.][]{TC2005}, are filtered out. Finally, it should be underlined, as this has been pointed out by \cite{Osborn1980}, that when shear instabilities set up, it could be difficult to disentangle fluctuating turbulent flows from wave motions \citep{AMD2013,Herbertetal2016,Lametal2021}.\\

\subsection{Impact of the Coriolis acceleration}

A key interesting point concerning the Osborn-Cox balance is that it has been derived with taking into account the Coriolis acceleration in the initial momentum equations. As a consequence, the related effective diffusivity can be applied to the vertical shear of low-frequency internal gravity and gravito-inertial waves. This result is coherent with the results obtained by \cite{Vargheseetal2025} where the measured effective diffusivity scales as in Eq. (\ref{Eq:Dstar}) independently of the rotation rate. This is also in agreement with the theoretical results obtained by \cite{ParkMathis2025}, where we showed that the inflectional instability of vertically sheared flows (here, the horizontal velocity of low-frequency internal gravity and gravito-inertial waves) is only weakly directly influenced by the Coriolis acceleration. Such a weak dependence has also been observed in nonlinear hydrodynamical equatorial simulations computed by \cite{ChangGaraud2021}.

\section{Discussion and conclusions}
\label{sec:DiscConclu}

In this article, we examine the recent results obtained on the mixing of chemicals in stably-stratified rotating stellar radiation zones using the so-called Osborn-Cox closure model \citep{OsbornCox1972,Osborn1980}, which is widely used in physical oceanography. This model splits flows in mean and fluctuating components in stably stratified rotating oceanic layers and proposes a parameterisation of the mass transport (the dyapicnal mixing) by an effective eddy-like diffusion for the density, the entropy, the temperature and the chemical composition \citep{Hamiltonetal1989}. Assuming that the fluctuating velocity field in stellar radiation zones is composed by low-frequency progressive internal gravity and gravito-inertial waves, the Osborn-Cox closure model leads to an effective eddy-like diffusion for the chemicals they transport that scales as the ratio of the product of the heat diffusivity multiplied by the squared vertical shear of the horizontal component of their velocity divided by the squared Brunt-V\"as\"al\"a frequency. This result is of high interest, since this effective diffusion with such dependences has been measured in state-of-the-art nonlinear hydrodynamical equatorial numerical simulations computed by \cite{Vargheseetal2025} for early-type and late-type stars at different rotation rates \citep[we refer also the reader to][]{RME2017,Vargheseetal2023,Vargheseetal2024}. Therefore, we obtain a complementary physical understanding of this measured effective diffusivity by showing that it is associated to a balance between the power transferred from the mean flow to the fluctuating velocities, their energy dissipation, and the power of their buoyancy flux that leads to mixing. This balance has been derived from the Navier-Stokes equations with taking simultaneously into account the buoyancy force and the Coriolis acceleration. Therefore, the obtained parameterisation can be applied both to internal gravity waves \citep[e.g.][]{Zahnetal1997,Mathis2025} and to gravito-inertial waves \citep[e.g.][]{Mathisetal2008,Mathis2009,Vargheseetal2024,Mathis2026a} in any rotating star along its evolution. This strenghtens the interest of its implementation both in existing 1-D and in forthcoming multi-D stellar structure and evolution models \citep[e.g.][]{TC2005,Mombargetal2025}.

However, this work does not close the quest for parameterisations of the wave-driven mixing. Indeed, such a mixing can be driven by other mechanisms such as the irreversible Stokes displacement \citep[e.g.][]{Press1981,Schatzman1993,MaoLecoanet2025} or the wave convective and shear-induced breaking \citep{Lindzen1981,Liuetal2025} with possible couplings of the different instabilities of the waves \citep[e.g.][]{LombardRiley1996,Howlandetal2021}. Therefore, our goal is to provide coherent parameterisations for each possible physical configuration.

\begin{acknowledgements}
S.M. warmly thanks the anonymous referee for their constructive suggestions, which have allowed to improve the article. He also acknowledges support from the European Research Council (ERC) under the Horizon Europe programme (Synergy Grant agreement 101071505: 4D-STAR), from the CNES SOHO-GOLF and PLATO grants at CEA-DAp, and from PNPS (CNRS/INSU). While partially funded by the European Union, views and opinions expressed are however those of the author only and do not necessarily reflect those of the European Union or the European Research Council. Neither the European Union nor the granting authority can be held responsible for them.
\end{acknowledgements}

\bibliographystyle{aa}  
\bibliography{Mathis2026} 

\begin{appendix}

\section{Explicit expressions for the different geometries}
\label{Appen:A}

In Cartesian coordinates in \cite{Mathis2026a}, $X_{h}=\chi$, the reduced horizontal coordinate introduced by \cite{GerkemaShrira2005}, ${\mathcal H}\left(X_{h}\right)\equiv\exp\left[i k_{\perp} \chi\right]$, and $\epsilon\left(r\right)=\left[k_{r}\left(r\right)\right]^{-1/2}\left[k_{r}\left(r\right)/k_{\perp}\right]\left[\left(f^2+{\widehat\omega}^2\right)/{\widehat\omega}^2\right]^{1/2}$, where $k_{r}$ is given by their Eq. (15), $f=2\Omega\cos\theta$ with $\Omega$ the angular velocity and $\theta$ the co-latitude, and ${\widehat\omega}\left(r\right)$ the wave's Doppler-shifted frequency.\\

In polar coordinates in \cite{Vargheseetal2024},  $X_{h}\equiv\phi$ the polar angle, ${\mathcal H}\left(X_{h}\right)\equiv\exp\left[i m \phi\right]$, and $\epsilon\left(r\right)={\overline\rho}^{\,-1/2}r^{-3/2}\left[k_{r}\left(r\right)\right]^{-1/2}\left[k_{r}\left(r\right)/k_{h}\left(r\right)\right]$, where $k_{r}=(N/{\widehat\omega})\,k_{h}$ with $k_{h}\equiv m/r$, where $m$ is the azimuthal degree.\\

In spherical coordinates, $X_{h}\equiv\left\{\theta,\varphi\right\}$, the colatitude and the azimuth, respectively, ${\mathcal H}\left(X_{h}\right)\equiv Y_{l}^{m}\left(\theta,\varphi\right)$ the usual spherical harmonics, and $\epsilon\left(r\right)={\overline\rho}^{\,-1/2}r^{-2}\left[k_{r}\left(r\right)\right]^{-1/2}\left[k_{r}\left(r\right)/k_{h}\left(r\right)\right]$, where $k_{r}=(N/{\widehat\omega})\,k_{h}$ with $k_{h}\equiv\sqrt{l\left(l+1\right)}/r$. 

Note that the Coriolis acceleration can be taken into account using the Traditional Approximation of Rotation if $2\Omega\!<\!\!<\!N$ \citep[e.g.][]{Mathisetal2008}. In this case the spherical harmonics can be replaced by the so-called Hough functions as well as their horizontal eigenvalues.

\end{appendix}

\end{document}